\documentclass[11pt]{article}
\usepackage{xspace}
\usepackage{graphicx}
\usepackage{amsmath}
\usepackage{amssymb}
\usepackage{color}
\usepackage{multirow}

\definecolor{Red}{rgb}{1,0,0}
\definecolor{Green}{rgb}{0,1,0}
\definecolor{Blue}{rgb}{0,0,1}
\definecolor{Black}{rgb}{0,0,0}

\def\beq{\begin{equation}}
\def\eeq#1{\label{#1}\end{equation}}
\def\eeqn{\end{equation}}

\def\beqa{\begin{eqnarray}}
\def\eeqa#1{\label{#1}\end{eqnarray}}
\def\eeqan{\end{eqnarray}}

\let\bar=\overbar

\def\Dslash{\not{\hbox{\kern-4pt $D$}}}
\def\dslash{\not{\hbox{\kern-2pt $\del$}}}

\def\msb{{\bar{\ssstyle M \kern -1pt S}}}

\def\Title#1{\begin{center} {\Large {\bf #1} } \end{center}}

\begin{document}

\Title{Physics Potential of an Advanced Scintillation Detector }
\begin{center}
{\Large Introducing \textsc{Theia} }
\end{center}
\bigskip\bigskip

\begin{raggedright}  

{\it Gabriel D. Orebi Gann\index{Orebi Gann, Gabriel D.} (on behalf of the \textsc{Theia} Interest Group)\\
Department of Physics, University of California, Berkeley, CA 94720, USA \\
Nuclear Science Division, Lawrence Berkeley National Laboratory, CA 94720, USA}

\end{raggedright}
\vspace{1.cm}

{\small
\begin{flushleft}
\emph{To appear in the proceedings of the Prospects in Neutrino Physics Conference, 15 -- 17 December, 2014, held at Queen Mary University of London, UK.}
\end{flushleft}
}

\section{Introduction}
The recent development of water-based liquid scintillator and the concurrent development of high-efficiency and high-precision-timing light sensors has opened up the possibility for a new kind of large-scale detector capable of a very broad program of physics.   The program would span topics in nuclear, high-energy, and astrophysics, ranging from a next-generation neutrinoless double beta decay search capable of covering the inverted hierarchy region of phase space, to supernova neutrino detection, nucleon decay searches, and measurement of the neutrino mass hierarchy and CP violating phase.  This paper describes the technical breakthroughs that led to this possibility, and the broad physics program thus enabled.  This paper is a summary of a talk presented at the NuPhys 2014 conference in London.

\section{The Advanced Scintillation Detector Concept}
The Advanced Scintillation Detector Concept (ASDC)~\cite{asdc} leverages a tried and tested methodology in combination with novel, cutting-edge technology.  
The future of neutrino detection technology lies in massive, high-precision detectors, offering multiple channels for detection.  Current technology is constrained by the choice of target material: water detectors are limited in energy threshold and resolution by the overall light yield of the Cherenkov process, and scintillator detectors are limited in size by optical attenuation in the target itself, and in reconstruction of event direction by the isotropic nature of scintillation light.  

The newly-developed water-based liquid scintillator (WbLS)~\cite{wbls} offers a unique combination of  high light yield and low-threshold detection with attenuation close to that of pure water, particularly at wavelengths $>$ 400~nm.  
Use of this novel target material could allow separation of prompt, directional Cherenkov light from the more abundant, isotropic, delayed scintillation light.  This would be a huge leap forwards in neutrino detection technology, enabling the first low-threshold, directional neutrino detector.  Such a detector could achieve fantastic background rejection using directionality, event topology, and particle ID.  
WbLS chemistry also allows loading of metallic ions as an additional target for
particle detection, including: $^7$Li for charged-current solar neutrino detection; $^{\rm nat}$Gd
for neutron tagging enhancement; 
or isotopes that undergo double beta decay, facilitating a
neutrinoless double-beta decay (NLDBD) program.  The formula and principle of mass-produced WbLS have been developed and demonstrated at the Brookhaven National Laboratory Liquid Scintillator Development Facility.  Metal-doped samples have been produced with high stability, with loadings of up to several percent.  The instrumentation for large-scale liquid production is currently under design.

\textsc{Theia} is a proposed realization of the Advanced Scintillation Detector Concept~\cite{asdc}, which combines the use of a 30--100-kton WbLS target, doping with a number of potential isotopes, high efficiency and ultra-fast
timing photosensors, and a deep underground location.  A potential site is the Long Baseline Neutrino Facility (LBNF) far site, where \textsc{Theia} could operate in conjunction with the liquid argon tracking detector proposed by DUNE~\cite{dune}.  
The basic elements of this detector are being developed now in experiments such as WATCHMAN~\cite{wm} and SNO+~\cite{sno+}.  
\textsc{Theia} would address a broad program of physics, including: solar neutrinos, geo-neutrinos, supernova neutrinos, 
nucleon decay, measurement of the neutrino mass hierarchy and CP violating phase, 
and even a next-generation NLDBD search.   
	
\section{Physics Program}

A large-scale WbLS detector such as \textsc{Theia} can achieve an impressively broad program of physics topics, with enhanced sensitivity beyond that of previous detectors.  Much of the program hinges on the capability to separate prompt Cherenkov light from delayed scintillation.  This separation provides many key benefits, including:
\begin{itemize}
\item {\it The potential to perform ring-imaging as in a pure water Cherenkov detector} (WCD).  This enables a long-baseline program in a scintillation-based detector.

\item {\it Direction reconstruction using the prompt Cherenkov photons}.   This allows statistical identification of events such as solar neutrinos, which form a background to many rare-event searches, including NLDBD and nucleon decay.
\item {\it Detection of sub-Cherenkov threshold scintillation light}.  This provides excellent particle identification, including enhanced neutron tagging, detection of sub-Cherenkov threshold particles such as kaons in nucleon decay searches, and separation of atmospheric neutrino-induced neutral current backgrounds in inverse beta decay searches.
\end{itemize}

One of the most powerful aspects of \textsc{Theia} is the flexibility: both in the target medium itself, and even in the detector configuration.  The WbLS target can be tuned to meet the most critical physics goals at the time by modifying features of the target cocktail, including: the fraction of water vs scintillator; the choice of wavelength shifters and secondary fluors; and the choice of loaded isotope.  There is also the potential to construct a bag to contain isotope, and perhaps a higher scintillator-fraction target, in the centre of the detector, building on work by KamLAND-Zen~\cite{klz} and Borexino~\cite{bor}.  
The following provides a summary of some of the potential capabilities of a detector like \textsc{Theia}.  More detail on each topic can be found in the ASDC concept paper~\cite{asdc}.  

\subsection{Long-Baseline Physics}
A large-scale WbLS detector underground at the LBNF far site at the Homestake mine in South Dakota would provide a secondary target for the high-energy neutrino beam directed towards the mine from Fermilab.  This would provide a complementary program to the LArTPC proposed by the DUNE collaboration~\cite{dune}, including independent checks of systematics, such as the interactions of neutrinos on LAr at GeV energy scales, as well as a broad program of additional physics topics.  \textsc{Theia} can build on extensive feasibility studies already completed for the original WCD~\cite{lbne}.

\begin{figure}[!ht]
\begin{center}
\includegraphics[width=0.48\columnwidth]{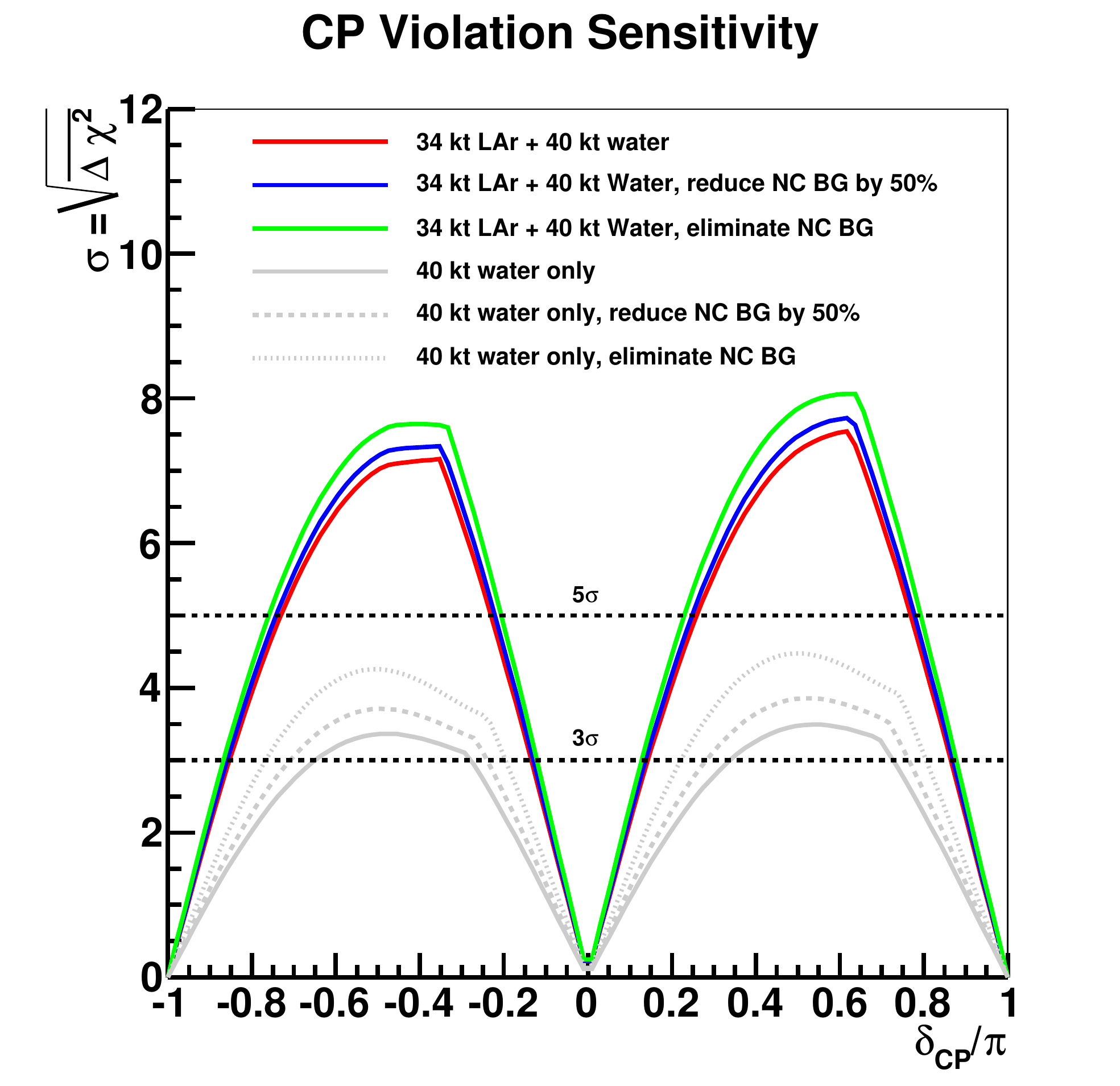}
\includegraphics[width=0.48\columnwidth]{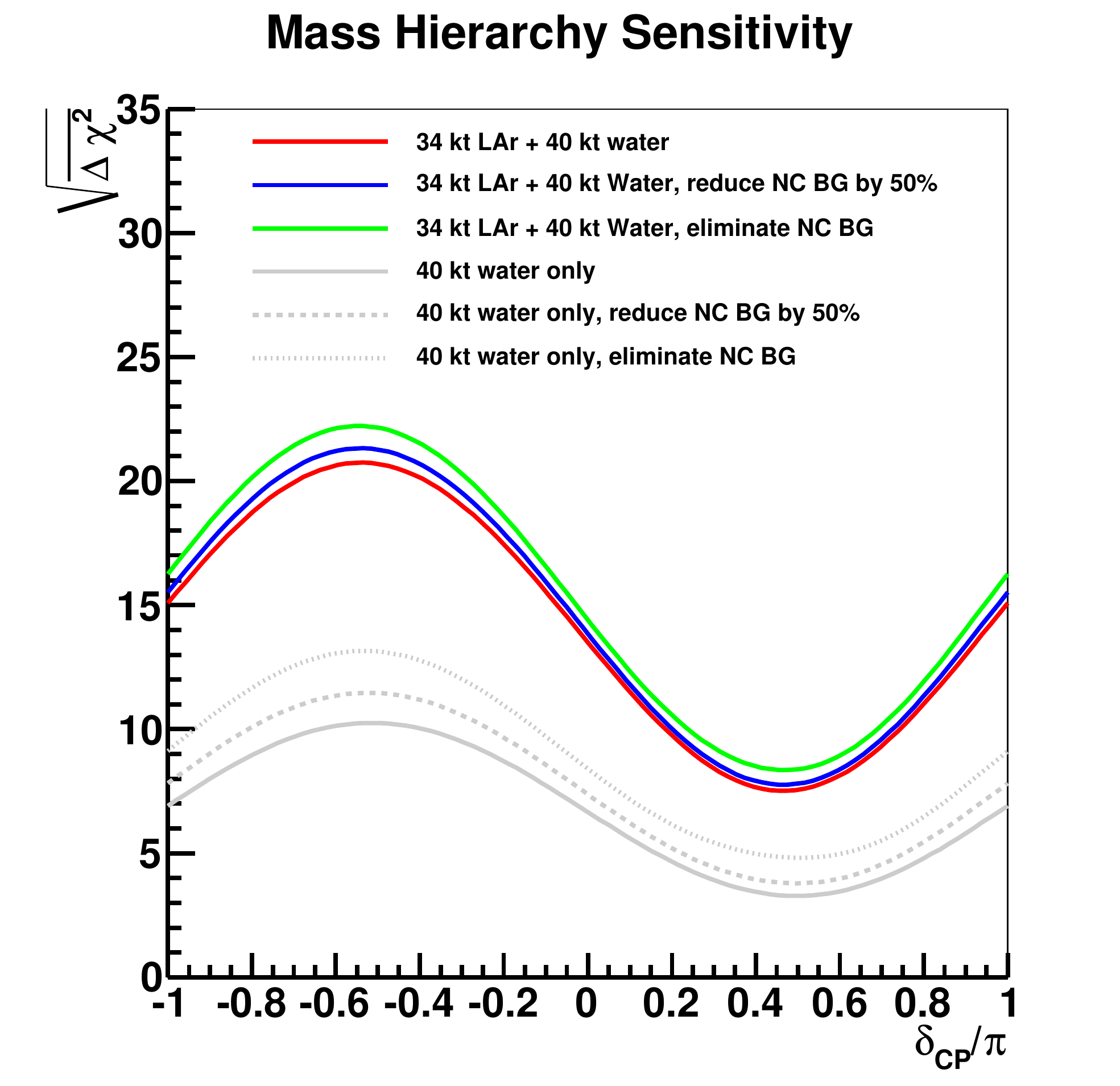}
\vspace{-0.25cm}
\caption{Significance of sensitivity to CP violation (left) and neutrino mass hierarchy (right), as a function
of the true value of $\delta_{CP}$, for a 34-kton LAr TPC in combination with an additional 40-kton WCD when neutral
current background is reduced by 50\% (blue) or removed (green) from the WCD event sample. Gray curves
show the sensitivity of the 40-kton WCD detector alone. All detector masses are fiducial and all sensitivities
are calculated at a baseline of 1300~km in the nominal LBNF beam. The normal hierarchy is assumed, and
oscillation parameters and uncertainties are taken from a recent global fit~\cite{globalfit}. The absolute sensitivity is
dependent on these parameters, in particular the choice of $\theta_{23}$, but the relative comparison is unaffected.  Figures taken from~\cite{asdc}.}
\vspace{-0.5cm}
\label{f:lbl}
\end{center}
\end{figure}

Cherenkov / scintillation separation in  \textsc{Theia} would allow the ring-imaging of a pure WCD.  Although some fraction of the Cherenkov photons would be absorbed and reemitted isotropically by the scintillator component, a comparison of Super-Kamiokande phase I and phase II data~\cite{lbne} illustrates that there is no loss in beam physics sensitivity even with a factor of two reduction in light yield.   \textsc{Theia} would also offer several enhancements beyond a WCD: detection of sub-Cherenkov threshold scintillation light allows discrimination between low-energy hadrons and electrons; the low-threshold scintillation light enhances neutron tagging efficiency, which allows statistical separation of neutrino and anti-neutrino events; and high-precision timing would improve reconstruction.  Together these all contribute to reduce the dominant neutral current (NC) background, which limits the sensitivity of a WCD.  Fig.~\ref{f:lbl} shows the improvement gained by this reduction in NC events.  The large scale of the detector also offers the potential to observe events at the second oscillation maximum, which could enhance the sensitivity to $\delta_{CP}$.

\subsection{Neutrinoless Double Beta Decay}
The search for neutrinoless double beta decay (NLDBD) has the potential to answer one of the most pressing open questions in neutrino physics today: the very nature of the neutrino.  Observation of this process would unequivocally demonstrate lepton number violation, and show that the neutrino is Majorana in nature.  One method by which to search for this process is to load a large-scale liquid scintillator (LS) detector with an isotope known to undergo two-neutrino DBD.  R\&D for the SNO+ NLDBD experiment has demonstrated the capability of loading percent levels of $^{130}$Te into WbLS with acceptable light levels~\cite{sno+} and further R\&D is expected to improve the optics considerably.  

SNO+ is a kton-scale detector, which plans to deploy a WbLS cocktail with a high fraction ($\approx 95\%$) of LS.  This provides high intrinsic light yield (\# optical photons generated per MeV of deposited particle energy), but also relatively high absorption and scattering in the target.  \textsc{Theia} is considering a much reduced LS fraction ($1-10\%$).  The  significantly reduced attenuation, particularly important as detector size increases, combined with high-efficiency, high-coverage photon detection can potentially result in an overall higher light collection (\# detected photoelectrons per MeV of particle energy).  This is a critical factor in the overall detector energy resolution, which provides the only handle for reducing the large $2\nu\beta\beta$ background. 
Initial studies show that 160 pe/MeV, close to the overall light collection of the SNO+ detector, is achievable in a 50-kton detector with a 1\% WbLS target~\cite{asdc}.     

The large-scale LS approach to NLDBD has several advantages, including: coincident tagging of internal backgrounds; reduction of external background via fiducialization; a source in / source out comparison, allowing background measurement and subtraction; and a well-understood and modeled detector geometry.  The dominant background in SNO+  comes from the irreducible $^8$B solar neutrinos.   Directional sensitivity in \textsc{Theia} would allow this background  to be reduced significantly, with only a small loss in signal efficiency.  
A naive sensitivity calculation can be performed assuming  
a 50~kton water-based liquid scintillator detector with a
0.5\% loading of natural Te, and an energy resolution of $\sim$~5.5\% (from the
projected 160 pe/MeV light yield). We assume here that the only non-negligible
backgrounds are the $2\nu\beta\beta$ events and $^8$B neutrino events, with a
conservative 50\% reduction of the latter using the direction cuts described above. To
eliminate external backgrounds we 
require events to be more than 5.5~m from the PMTs, providing a total
fiducial mass of 30~kton. For  0.5\% loading this is  150~ton
of natural Te, or $\sim$ 50~ton of $^{130}$Te.  
 With this detector configuration, 
\textsc{Theia} could reach a 3$\sigma$ discovery of NLDBD in 10 years for
$m_{\beta\beta}=15$~meV~\cite{asdc}.

There is  potential to improve this significantly, potentially even pushing towards sensitivities in the normal hierarchy region~\cite{biller}, by deploying a higher fraction of isotope in a bag in the centre of the detector, thus at the same time increasing signal statistics and reducing uniform backgrounds such as $^8$B neutrinos.  Use of a higher scintillator fraction WbLS  would improve the energy resolution, reducing the $2\nu\beta\beta$ background.  Further studies are required to optimize the WbLS target for a NLDBD search in \textsc{Theia}, balancing energy resolution (reduction of $2\nu\beta\beta$ background) against directional resolution (reduction of $^8$B neutrinos).

\subsection{Solar Neutrinos}   

There are many open questions in solar neutrinos, several of which can be addressed by \textsc{Theia}'s combination of a low-threshold directional detector, along with the potential for isotope loading.  \textsc{Theia} would provide unprecedented sensitivity to solar neutrinos via two channels:
\begin{enumerate}
\item {\it Huge statistics for elastic scattering (ES) events at low energy}.  
The LENA collaboration~\cite{lena} have explored in detail the power of a large-scale scintillator detector for resolving open questions in solar neutrino physics, such as determining the solar metallicity via a measurement of neutrinos from the sub-dominant CNO fusion cycle.  \textsc{Theia} would have similar capability, along with the additional advantage of being able to distinguish ES events from backgrounds (such as $^{210}$Bi) using directionality.

\item {\it Potential charged-current (CC) detection via isotope loading e.g. $^7$Li}~\cite{li}.  
The differential CC cross section for neutrino interaction on $^7$Li is extremely sharply peaked.  As a result, CC neutrino detection provides a high-precision measurement of the incoming neutrino energy, allowing extraction of the low-energy $^8$B spectrum. This would provide a sensitive search for new physics via a probe of the transition region in the neutrino spectrum between vacuum-dominated and matter-enhanced oscillations.  There is also the potential to separate the different components of the CNO flux via a shape analysis.

\end{enumerate}

\subsection{Supernova Neutrinos}
\textsc{Theia} would observe roughly 12,000 events from a supernova at 10~kpc in a 50-kton fiducial volume.  90\% of these events would be detection of antineutrinos via inverse beta decay (IBD).  The low-threshold scintillation light in \textsc{Theia} would enhance the efficiency of the neutron tag, and this could be further improved by loading with Gadolinium, which would reduce the neutron capture time by roughly an order of magnitude, from 200~$\mu$s to 20~$\mu$s, thus reducing the risk of pile up from a large burst.  The high-efficiency tag for IBD events allows extraction of other event types from beneath this dominant signal.  Reduction of the IBD background for elastic scattering detection doubles the pointing accuracy, and identification of the CC and mono-energetic gammas from NC interactions provide sensitivity to the burst temperature and subsequent neutrino mixing.  It is worth noting that this antineutrino-dominated signal would be highly complementary to the neutrino-dominated signal in a liquid argon detector, such as DUNE.

\subsection{Diffuse Supernova Background Neutrinos}

The diffuse supernova neutrino background (DSNB) consists of both neutrinos and antineutrinos.  The antineutrino signal would be detected via IBD and, as for supernova burst neutrinos, \textsc{Theia}'s advantage over water Cherenkov detectors lies in the high efficiency neutron tag.  The low-threshold scintillation light provides excellent efficiency for detecting the 2.2~MeV gamma from neutron capture on $^1$H, effectively suppressing the singles rate that limits water Cherenkov detectors.  The efficiency would be greater even that Gd-loaded water detectors.  In comparison to a pure scintillator detector, \textsc{Theia} benefits from the use of the Cherenkov signal to enhance particle identification.  The primary background to a pure scintillator measurement is atmospheric neutrino induced NC reactions: the neutrinos strike carbon atoms, breaking it apart and causing the nuclear fragments to recoil, which is followed by capture of any liberated neutrons.   In a scintillator detector this coincidence mimics the IBD signal.  \textsc{Theia} could use the Cherenkov signal to discriminate between the nuclear recoil, which would be below Cherenkov threshold, and the positron of an IBD event.  The Cherenkov hit pattern would also be distinct for the two event types.

The possibility also exists to load \textsc{Theia} with NaCl in order to detect the DSNB neutrino signal via CC interaction on chlorine.

\subsection{Geoneutrinos}
Geoneutrinos are antineutrinos produced by radioactive decay inside the Earth.  Geoneutrino detection can help us to understand models for heat production beneath the surface of the Earth.  It is important to perform measurements in different geographical locations, due to the varying reactor neutrino background (an antineutrino signal with a similar energy spectrum), and in order to understand the different contributions from crust and mantle.  The total worldwide exposure is currently less than 10~kt-yr, from a combination of KamLAND~\cite{KLgeo} in Japan, and Borexino~\cite{Bgeo} in Italy.  \textsc{Theia} would provide a large statistics data set in a complementary geographical location, with the potential advantages of isotope loading and directionality, due to the fast timing and resulting high-precision reconstruction.

\subsection{Nucleon Decay}
A search for nucleon decay allows us to probe the possibility of grand unified theories.  \textsc{Theia}'s advantages lie in the depth and cleanliness of the detector, the large size, the high-efficiency neutron tag, and the ability to detect particles below Cherenkov threshold.

In the simplest case of heavy boson exchange we expect a lepton+meson final state.  The enhanced neutron tag would reduce the atmospheric neutrino background in \textsc{Theia}, allowing a measurement that would quickly become competitive with Super-Kamiokande, although Hyper-Kamiokande would dominate due to sheer size.  

Should the tree-level decay be suppressed, decay can proceed via modes such as $p\rightarrow~\nu~K^+$.  Here \textsc{Theia} has a strong advantage due to the ability to detect the sub-Cherenkov threshold kaon, leading to a triple coincidence tag.

More exotic theories, such as extra dimensions, predict so-called invisible modes like $n\rightarrow 3\nu$.  The low background at \textsc{Theia}, due to depth, cleanliness, and the ability to reject both solar and reactor events using directionality and the neutron tag, results in a sensitivity that dominates over other experiments.

More details on the sensitivity to different modes can be found in~\cite{asdc}.

\subsection{Sterile Neutrinos}

The existence of a fourth ``sterile'' neutrino has been suggested by several anomalous results in the neutrino sector, including LSND, MiniBooNE, and the reactor anomaly.  Many projects are proposed to search for sterile neutrinos.  \textsc{Theia} could perform a source-based search by deploying the $^8$Li decay-at-rest IsoDAR source~\cite{isodar}.  The 13~MeV endpoint puts the majority of these events well above radioactive backgrounds, and it places only conservative requirements on detector performance: 15\% energy and 50~cm position resolution.  Five years of data in a (conservative) 20-kton fiducial volume would exclude the majority of parameter space allowed by the various anomalies~\cite{asdc}.

\section{Required R\&D and Planned Demonstrations}

There are a number of demonstrations required in order to realize the conceptual detector presented here.  These include (but are not limited to): 
\begin{enumerate}
\item Sufficiently high intrinsic light yield and long attenuation length to meet minimal light collection requirements.   (This  requirement can be offset by high efficiency, high coverage photon detection).
\item Successful separation of  Cherenkov and scintillation signals, with sufficiently high Cherenkov light yield to maintain direction resolution and ring imaging capability.  This can be achieved by ultra-fast timing photon detection, such as LAPPDs~\cite{lappd2}, tuning of the WbLS cocktail, or a combination of the two.
\item Stability of the above properties over long timescales, and with respect to isotope loading ({\it e.g.} Gd, Li, Te).
\item Materials compatibility studies.
\item Demonstrated reconstruction \& particle ID capability.
\end{enumerate}

The R\&D program for \textsc{Theia} strongly leverages existing efforts.  A number of demonstrations are planned, ranging from bench-top to kton scale.  Table~\ref{tab:example} summarizes the status, including scale, target, timescale, and the measurements planned at each site.

\begin{table}[!th]
\begin{center}
\caption{Planned demonstrations of WbLS and fast-timing technology.}
\begin{tabular}{lcc p{5cm} c}  \hline\hline
Site & Scale & Target & Measurements & Timescale \\
  \hline
UChicago		&bench top	&	\multirow{2}{*}{H$_2$O}		&Fast photodetectors.		& Exists \\
CHIPS		& 10kton		&							& Electronics, readout, mechanical infrastructure. & 2019 \\
\hline
 EGADS		& 200 ton 		&	\multirow{3}{*}{H$_2$O+Gd} 	&\multirow{3}{*}{Loading, fast photodetectors.}			&	Exists \\
ANNIE		&1 ton		&							& 			& 2016 \\
WATCHMAN	& 1 kton		&							&			& 2019 \\
\hline
UCLA/MIT		&	1 ton		&	LS						&	Fast photodetectors.	&2015\\
\hline
Penn			&	30 L		&	\multirow{2}{*}{(Wb)LS}	&	\multirow{2}{*}{Light yield, timing, loading.} &Exists \\
SNO+		&780 ton		&						&			&2016	\\
\hline
UC Irvine		& 10 	L		&   \multirow{4}{*}{WbLS} 		& Attenuation, recirculation & 2015 \\
LBNL		&bench top	&		&	\multirow{3}{5cm}{Light yield, timing, cocktail optimization, loading, attenuation, reconstruction.} 	& 2015\\
BNL			&1 ton		&		&		& 2015\\
WATCHMAN-II	&1 kton		&		&		&2021\\
  \hline\hline
\end{tabular}
\label{tab:example}
\end{center}
\end{table}

\section{Summary}

\textsc{Theia} represents potentially revolutionary technology.  
Use of the novel, potentially inexpensive WbLS target  allows construction of a precision detector on a massive scale.  
Successful identification of Cherenkov light in a scintillating detector would result in unprecedented background-rejection capability and signal detection efficiency via directionality and sub-Cherenkov threshold particle identification.  This low-threshold, directional detector could achieve a fantastically broad physics program, combining conventional neutrino physics with rare-event searches in a single, large-scale detector.

\textsc{Theia} offers a broad program of compelling science, covering topics in nuclear physics, high-energy physics, astrophysics, and geophysics.  These include: solar neutrinos and neutrinoless double beta decay; proton decay and long-baseline physics; supernova neutrinos and DSNB;  and geoneutrinos, respectively.

The flexibility of the WbLS target, of the options for isotope loading, and even of the detector configuration is a crucial aspect of \textsc{Theia}'s design.  The status of the field will evolve during the planning and realization of a project of this scale; \textsc{Theia} has the unique ability to adapt to new directions in the scientific program as the field evolves, making it a powerful instrument of discovery that could transform the next-generation of experiments.

\bigskip
\section{Acknowledgments}

This material is based upon work supported in part by the U.S. Department of Energy, Office of Science, Office of High Energy Physics, the Department of Energy
National Nuclear Security Administration under Award Number: DE-NA0000979
through the Nuclear Science and Security Consortium.  
Research conducted at the Brookhaven National Laboratory is sponsored by the Office of Nuclear Physics, Office of High Energy Physics, Office of Science, United States Department of Energy under contract DE-SC0012704, and by the Laboratory Directed Research and Development Program of Brookhaven National Laboratory, No. 12-033.  
Research conducted at Lawrence Berkeley National Laboratory is supported by the Laboratory Directed Research and Development Program of Lawrence Berkeley National Laboratory under U.S. Department of Energy Contract No. DE-AC02-05CH11231, and the University of California, Berkeley.  
Research conducted at Lawrence Livermore National Laboratory is supported by the LDRD Program, Tracking Number 15-ERD-021.


\begin{thebibliography}{99}

\bibitem{asdc} J. R. Alonso {et al.},
arXiv: 1409.5864 [hep-ex, nucl-ex]


\bibitem{wbls}M. Yeh {\it et al.}, 
Nucl. Inst. \& Meth. A{\bf 660} 51 (2011)

\bibitem{dune} C. Adams {et al.},
arXiv: 1307.7335 (2013)

\bibitem{wm} M. Askins  {et al.},
arXiv: 1502.01132 (2015)

\bibitem{sno+} SNO+ Collaboration, {\it White Paper, under development}; M. C. Chen for the SNO+ Collaboration, arXiv: 0810.3694 (2008)

\bibitem{klz} KamLAND-Zen Collaboration, Phys.\ Rev.\ Lett.\  {\bf 110}:062502 (2013)

\bibitem{bor} G. Bellini {\emph et al.}, Phys. Rev. Lett. {\bf 107} 141302 (2011); G. Bellini {\it et al.},  Phys. Rev. Lett. {\bf 108} 051302 (2012);  G. Bellini {\it et al.}, Nature {\bf 512} 383-386 (2014)

\bibitem{lbne} The LBNE Collaboration, 
arXiv: 1204.2295 (2012)

\bibitem{globalfit} F. Capozzi  {et al.},
Phys.\ Rev.\ D  {\bf 89} 093018 (2014) 

\bibitem{biller} S. D. Biller, 
Phys. Rev. D {\bf 87} no. 7 : 071301 (2013)  

\bibitem{lena} M. Wurm  {\it et al.}, arXiv:1104.5620 [astro-ph.IM] (2011)

\bibitem{li}W. Haxon,  Phys. Rev. Lett.{\bf 76} 10 (1996)

\bibitem{KLgeo} A. Gando \emph{et al.}, Phys. Rev. \textbf{D88} 033001 (2013)
\bibitem{Bgeo} G. Bellini \emph{et al.}, Phys. Lett. \textbf{B722} 295-300 (2013)

\bibitem{isodar} A. Bungau \emph{et al.}, Phys.\ Rev.\ Lett.\ {\bf 109}, 141802 (2012)

\bibitem{lappd2}O.H.W. Siegmund {\it et al.}, 
JINST {\bf 9} (2014) C04002

\end{thebibliography}
\end{document}